# Test suite effectiveness metric evaluation: what do we know and what should we do?

Peng Zhang  Yang Wang  Xutong Liu  Yibiao Yang
Yanhui Li  Lin Chen  Ziyuan Wang  Chang-ai Sun  Yuming Zhou

National Key Laboratory for Novel Software Technology, Nanjing University
School of Computer Science, Nanjing University of Posts and Telecommunications
School of Computer & Communication Engineering, University of Science and Technology Beijing

## ABSTRACT

Comparing test suite effectiveness metrics has always been a research hotspot. However, prior studies have different conclusions or even contradict each other for comparing different test suite effectiveness metrics. The problem we found most troubling to our community is that researchers tend to oversimplify the description of the ground truth they use. For example, a common expression is that "we studied the correlation between real faults and the metric to evaluate (MTE)". However, the meaning of "real faults" is not clear-cut. As a result, there is a need to scrutinize the meaning of "real faults". Without this, it will be half-knowledgeable with the conclusions. To tackle this challenge, we propose a framework ASSENT (evAluating teSt Suite EffectiveNess meTrics) to guide the follow-up research. In nature, ASSENT consists of three fundamental components: ground truth, benchmark test suites, and agreement indicator. First, materialize the ground truth for determining the real order in effectiveness among test suites. Second, generate a set of benchmark test suites and derive their ground truth order in effectiveness. Third, for the benchmark test suites, generate the MTE order in effectiveness by the metric to evaluate (MTE). Finally, calculate the agreement indicator between the two orders. Under ASSENT, we are able to compare the accuracy of different test suite effectiveness metrics. We apply ASSENT to evaluate representative test suite effectiveness metrics, including mutation score metrics and code coverage metrics. Our results show that, based on the real faults, mutation score and subsuming mutation score are the best metrics to quantify test suite effectiveness. Meanwhile, by using mutants instead of real faults, MTEs will be overestimated by more than 20% in values.

## 1  Introduction

Knowing the effectiveness of a test suite in fault detection is one of the most fundamental tasks in software testing [1-4]. To this end, many technologies have been proposed to evaluate the effectiveness of test suites. Coverage testing and mutation testing are the two most widely used technologies. For coverage testing, a set of predefined covering requirements are first monitored when executing the software under test (SUT) against a test suite. Then, code coverage is computed for the suite to measure its effectiveness. For mutation testing [5], a set of predefined mutators are first used to simulate the possible faults for the SUT, which are referred as mutants. Then, the test suite is executed to determine which mutants are killed (i.e., detected). After that, a mutation score is obtained by computing the ratio of killed mutants to measure the test suite effectiveness.

By the common sense of existing studies, mutation score is regarded as a more accurate metric than code coverage to estimate the effectiveness of a test suite [6, 7]. However, mutation testing is much more expensive than coverage testing as it requires executing the test suite for many more times than the coverage testing to check each of those mutants. As a result, there are two main practices for improving the two technologies. For coverage testing, many fine-grained coverage criteria have been proposed for more accurate effectiveness evaluation [8-10]. For mutation testing, how to reduce the execution overhead, even if it brings the loss of accuracy, has become a research hotspot [11-29]. Regardless of their specific purpose, in nature, a practice on coverage testing provides a code coverage metric (e.g., branch coverage) while a practice on mutation testing provides a mutation score metric.

A natural question is how to evaluate the accuracy of a test suite effectiveness metric. The most common practice is to perform correlation analysis on the independent variables (i.e., MTEs) and dependent variable (i.e., ground truth). However, the biggest pitfall is that the description of the ground truth is too simplified. For example, in [41], the meaning of "fault detection" as the ground truth is not clear enough. By "fault detection", our intuition is that a high fault detection represents high effectiveness. However, after carefully scrutinizing the ground truth used in [41], we find that the authors use "fault detection" to mean that *given a number of test suites with the same size and a specific fault, the test suites that can detect this fault are more effective than the test suites that cannot detect this fault*. Until now, the threat of the ground truth has come to the surface. For instance, based on this ground truth, we can easily deduce that the triggering test case is more effective than any other test case. Obviously, this assumption deserves further verification. By this example, we reveal that an oversimplified expression of ground truth makes the reader to make intuitive deductions that may not be consistent with what the author intended to convey. To address this issue, this paper attempts to propose an evaluation framework including fully articulated ground truth to help researchers to have a more intelligible evaluation of different effectiveness metrics.

**Table 1. Test effectiveness metric evaluation studies.** The "ground truth" is when they assume A is more effective than B. "FD" is the number of faults detected by the corresponding test suite. "MS" is the mutation score of the corresponding test suite. "OP" is explained in Section 3.3.

| Paper | Year | Venue | Ground Truth | Benchmark Test Suites | Agreement Indicator | Summary of Scientific Findings |
|---|---|---|---|---|---|---|
| Chen et al. [39] | 2020 | ASE | FD(A)>FD(B) | Greedily adding one test into B to achieve a higher adequacy for a MTE to obtain A | - | Adequacy-based test selection is superior to random selection. Mutation-based test selection is most effective when employed after coverage has exhausted its usefulness |
| Hariri et al. [40] | 2019 | ICST | $|A|=|B|$, FD(A)>FD(B) | T is the test pool. A,B are randomly sampled of fixed size from T. | Kendall Pearson | There is a very weak correlation between mutant detection ratio and real fault detection ratio. |
| Papadakis et al.[41] | 2018 | ICSE | $|A|=|B|$, FD(A)>FD(B) | T is the test pool. A,B are randomly sampled of fixed size from T. | Kendall Pearson | There is a weak correlation between mutant detection ratio and real fault detection ratio. |
| Shin et al. [42] | 2018 | TSE | FD(A)>FD(B) | A,B are generated by keeping greedily selecting test cases until the MTE value is in a given interval. | Rank-biserial Correlation | Distinguishing mutation adequacy criterion has higher fault detection probability than strong mutation adequacy criterion. |
| Chekam et al. [43] | 2017 | ICSE | $|A|=|B|$, FD(A)>FD(B) | T is the test pool. A,B are randomly sampled of fixed size from T. | - | There is a strong connection between mutation score increase and fault detection at higher score levels. |
| Gligoric et al. [30] | 2015 | TOSEM | $|A|=|B|$, MS(A)>MS(B) | A,B are generated by keeping randomly selecting test cases until the MTE value is in a given interval. | Kendall Spearman R | There is a strong correlation between mutant detection ratio and branch coverage. |
| Kochhar et al. [4] | 2015 | SANER | $|A|=|B|$, FD(A)>FD(B) | T is the test pool. A,B are randomly sampled of fixed size from T. | Point Biserial Correlation | There is a moderate to strong correlation between mutant detection ratio and real fault detection ratio. |
| Gopinath et al. [31] | 2014 | ICSE | MS(A)>MS(B) | Each test set (developer-written and automatically-generated) was used for analysis without sampling. | Kendall R | There is a strong correlation between mutant detection ratio and statement coverage. |
| Inozemtseva and Holmes [33] | 2014 | ICSE | $|A|=|B|$, MS(A)>MS(B) | T is the test pool. A,B are randomly sampled of fixed size from T. | Kendall Pearson | There is a negligible to moderate correlation between mutant detection ratio and statement coverage. |
| Just et al. [34] | 2014 | FSE | $A \approx B$, FD(A)>FD(B) | B passes on buggy and fixed version A fails on buggy version but passes on fixed version A and B differ by one modified or added test | OP | There is a strong correlation between mutant detection ratio and real fault detection ratio. |
| Namin et al. [45] | 2009 | ISSTA | $|A|=|B|$, MS(A)>MS(B) | T is the test pool. A,B are randomly sampled of fixed size from T. | Kendall Pearson R | For samll suites, there is a moderate to very high correlation between mutant detection ratios and coverage. For large test suites, there is a low to moderate correlation. |
| This paper | | | $B \subsetneq A$, FD(A)>FD(B) | A fails on buggy version but passes on fixed version. B is obtained by deleting triggering test cases from A. | OP | Mutation score and subsuming mutation score are the best metrics |

Most importantly, to have a fair comparison for different test suite effectiveness metrics, researchers need to determine a ground truth. The ground truth is used to answer the question: when can we say that test suite A is more effective than test suite B in defect detection? For example, if A is the test suite that contains triggering test cases while B is the subset of A by removing those triggering test cases from A. We can say that A is more effective than B. This is because B is a subset of A, and A can detect a real defect while B cannot. After determining the ground truth, researchers need to generate comparable A and B. Then, by the metric to evaluate (MTE), we reevaluate the test effectiveness for the set of test suites. Finally, the agreement between the two evaluations is calculated. For example, in [30], when comparing several coverage metrics using mutation score as the ground truth, the correlation coefficients between coverage and mutation score are calculated. In their study, a higher coefficient means the coverage metric is better.

To summarize, an evaluation process has to answer the following three questions:

- When can we say that test suite A is more effective than test suite B? (ground truth)
- How to generate A and B? (benchmark test suites)
- Which indicator should be chosen to report the agreement between the MTE and the ground truth? (agreement indicator)

Table 1 lists the representative studies with "ground truth", "benchmark test suites" and "agreement indicator". Clearly, with such a framework, researchers can easily compare the reasons for different conclusions among multiple studies. This is one of our major motivations for proposing a standardized evaluation framework. From the table, we can see that the primary reason for the controversy in existing research is that different ground truths are used. As a result, we recommend that the question on ground truth should be answered explicitly in future work.

We named this framework ASSENT: ev**A**luating te**S**t **S**uite **E**ffective**N**ess me**T**rics. In this paper, we use ASSENT to evaluate the accuracy of seven test suite effectiveness metrics. First, we use the ability of different test suites in detecting real faults as the ground truth to determine the ground truth order among test suites. In particular, for a test suite A with triggering test cases for the specific fault, we filter all the triggering test cases to obtain a subset B. Then, we can say that A is more effective than B in term of the particular fault. This relation will be held not only for the buggy version but also for the fixed version. The reasons are two folded: for the buggy version, A can reveal the fault while B cannot; for the fixed version, the correctness is verified by the triggering tests cases. Second, for a buggy program, we use the developer-written tests with triggering test cases as the test suite A. Then, we filter all the triggering test cases to obtain B. Third, by a MTE, we use MTE to compare the effectiveness of A and B and to check whether the ground truth order is preserved. For instance, when evaluating statement coverage, if A achieves a higher coverage score than B, the ground truth order is preserved. If A achieves the same score compared to B, the ground truth order is twisted. After all faults are taken into account, the order preserving score will be calculated to evaluate estimate how accurate a MTE is in evaluating the effectiveness of test suites.

The contributions of our work include:
- We propose a practical framework ASSENT for comparing different test suite effectiveness metrics.
- We analyze the possible threats of those ground truths in existing studies and propose a new ground truth based on real defects.
- Based on ASSENT and the proposed ground truth, we compare the accuracy of a variety of representative test suite effectiveness metrics.

The rest of this paper is organized as follows. Section 2 introduces the background. Section 3 presents ASSENT. Section 4 describes the settings of experiments. Section 5 reports the experimental results. Section 6 discusses the experimental results. Section 7 analyzes the threats to the validity. Section 8 concludes the paper.

## 2  Background

In this section, we will provide an overview of the most commonly used testing practices for generating mutation score metrics, code coverage metrics, and related work in metric evaluation.

### 2.1  Mutation testing

In the literature, the following mutation reduction strategies are often used to reduce the number of mutants before a mutation score metric is computed.

*COS (Certain Operator Selection)* [11-16]: select the mutants generated by a specific mutator subset of all mutators to compute the mutation score. There are numerous practices on COS whose core idea is to remove redundant mutation operators and select important operators that are more related to defects. Mathur first proposed COS in 1991 and called it selective mutation [16]. They found that the number of redundant mutants could be reduced by 30% to 40% by deleting "arithmetic operator replacement" mutator and "scalar variable replacement" mutator. Offutt et al. found five important mutation operators for COS, including the relational, logical, arithmetic, absolute, and unary insertion operators [12].

*RMS (Random Mutant Selection)* [22-24]: randomly select a specified number or proportion of mutants from all mutants to compute the mutation score metric.

Beside mutant reduction, other practices aim at providing more accurate metrics. These metrics are performed after all mutants are executed.

*SMS (Subsuming Mutation Score)* [17-20]: select the subsuming mutants to compute the mutation score. Subsuming mutation score is proposed for avoiding inflation. By the definition in [20], "One mutant subsumes another if at least one test kills the first and every test that kills the first also kills the second". For all mutants, killing the subsuming mutants is to kill all the killable mutants. Therefore, the subsumed mutants should be regarded as redundant. The practical difficulty of SMS lies in how to figure out the "true" subsumption relationships among all the mutants. In [20], Kurtz et al. proposed both dynamic and static methods to approximate the "true" subsumption. In [18], Gong et al. manually inserted mutant branches into the original program to generate a giant program which was used to analyze for the approximation of the "true" subsumption.

*CMS (Clustering Mutation Score)* [21]: select one mutant from each mutant cluster to compute the mutation score. First, for all the mutants, a number of features are collected. Second, based on these features, a clustering algorithm is used to group the mutants into different clusters. Third, the selection is generated by selecting one mutant from each mutant cluster. For example, Hussain [21] executed all the mutants to obtain the information on whether a test can kill a mutant against all the tests and mutants. After that, K-means and agglomerative clustering were applied to cluster the mutants. By the clustering algorithm, the mutants in the same cluster were guaranteed to be killed by a similar set of test cases. Therefore, randomly selecting one mutant from each cluster can be seen as a selection of representative mutants.

### 2.2  Code coverage testing

Statement coverage and branch coverage provide the most common and simple code coverage metrics in practice. Recently, it has been shown that branch coverage is superior to statement coverage when comparing the effectiveness of test suites in defect detection [30]. In this paper, we are interested in comparing the accuracy of mutation score metrics and these two code coverage metrics in evaluating the effectiveness of test suites in defect detection.

### 2.3  Test effectiveness metric evaluation

Table 1 lists representative test effectiveness metric evaluation studies. From their findings, we can find that some of the conclusions are contradictory. For example, from [41], we know there is a weak correlation between mutation score and real fault detection, while from [34] there is a strong correlation between mutation score and real fault detection. In this paper, we point out that the fundamental elements leading to different results are "ground truth", "benchmark test suites" and "agreement indicator" (especially the ground truth).

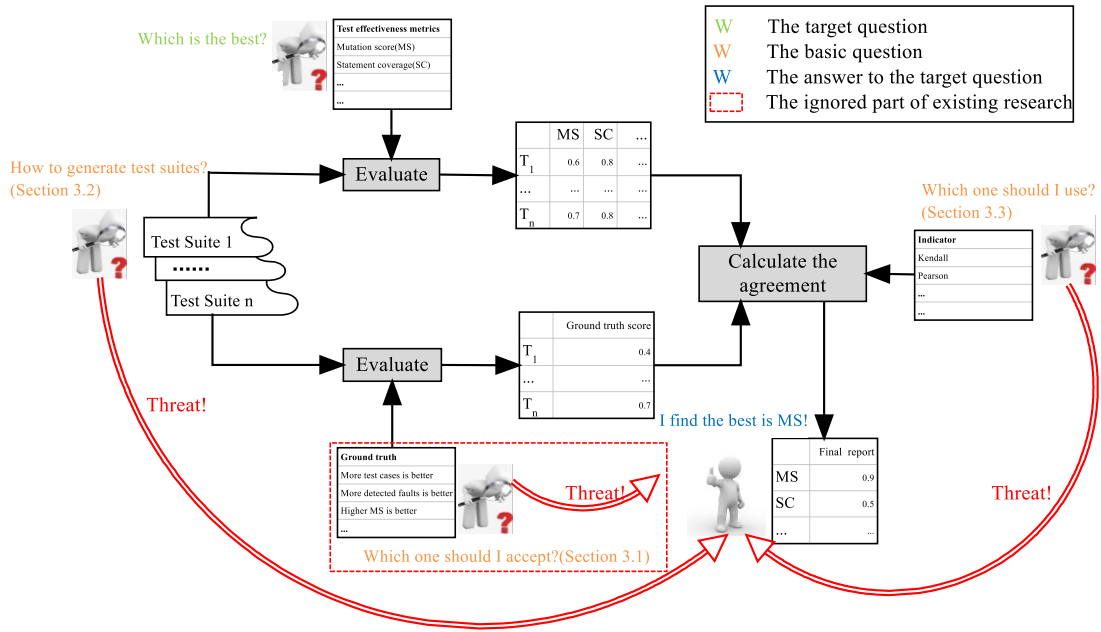

**Figure 1. ASSENT for the evaluation of test effectiveness metrics.** The "ground truth" is the ignored part of specific representative research. For example, in [41], there was no explicit answer to "which one should I accept". As a result, readers need to deduce the basis for judging the effectiveness according to the indicator they used (i.e., Kendall). On the one hand, the ignored part increases the difficulty of understanding. On the other hand, it makes the threat hidden deeper. A detailed analysis of this case is in Section 3.1.1.

Note that a few studies were not about correlation analysis. For example, in [39], they did not investigate the agreement between the MTE and the ground truth. They investigated which MTE is the most effective to select test cases. However, same to the mainstream, we will focus on the agreement.

## 3 ASSENT: evAluating teSt Suite EffectiveNess meTrics

The purpose of ASSENT is to evaluate the extent to which a given test effectiveness metric can replace the ground truth, i.e., agreement. The higher the agreement, the closer its evaluation of the test suites is to the evaluation under the ground truth.

There are three core elements in ASSENT. First, there should be a fully articulated ground truth. Second, there should be benchmark test suites to obtain the empirical relations under the ground truth on the one hand, and to calculate the numerical relations under the MTE (the metric to evaluate) on the other hand. Finally, there should be indicators to calculate the agreement between empirical and numerical relations. Fig. 1 shows the ASSENT for evaluating the accuracy of a test suite effectiveness metric.

### 3.1 Ground truth determination

Let us imagine a specific question as the motivation example to derive the empirical relationship: given two test suites A and B for a SUT, when can we say that A is more effective than B in defect detection? There are a few possible answers:

**Proposition 1:** If the size of A is bigger than B, A is deemed as more effective than B.

**Proposition 2:** If B is a proper subset of A, A is deemed as more effective than B.

**Proposition 3:** If A achieves a higher code coverage than B, A is deemed as more effective than B.

**Proposition 4:** If the covering requirement set covered by B is a proper subset of that by A, A is deemed as more effective than B.

**Proposition 5:** If A achieves a higher mutation score than B, A is deemed as more effective than B.

**Proposition 6:** If the mutant set detected by B is a proper subset of that by A, A is deemed as more effective than B.

**Proposition 7:** If the real fault set detected by B is a proper subset of that by A, A is deemed as more effective than B.

……

The purpose of this example is to make the ground truth criterion concrete rather than discussing these propositions. These propositions summarize the empirical knowledge of the relationship among test suites. By taking some propositions, we can design the ground truth criteria to derive the ground truth order among test suites. Note that this paper does not adopt exactly one proposition.

After taking a brief idea of the ground truth criterion, the following parts are two folded: first, we discuss the possible threat to the ground truth used in existing research; second, we propose the ground truth used in this paper.

*3.1.1 Threat to existing research.* In this subsection, we will analyze the possible threat to representative studies that comes from the ground truth. At the high level, the threat of using simulated faults (i.e., mutants) as the ground truth is obvious, which mainly comes from the difference between the mutants and real defects. However, there is still one question that has not been answered: will this approach lead to overestimation or underestimation of the MTE? The question is answered in our experiment. Therefore, in this subsection, we focus on the threat of using real faults as the ground truth.

In [41], researchers found that there is a weak correlation between mutation score and real fault detection when controlling the test suite size. For the test suites with the same size, they assumed that those which can detect the real fault are more effective than the others. This ground truth is implied by their indicator Kendall coefficient. It seems that they not only controlled the size, but also used real defects. As a result, their research looks "solid". However, we can use a simple example to show its pitfall of "size". Assume that for a fault, there is only one triggering test case t in the test pool T. Then we let A={t} and B={t'} where t' ∈T/{t}. By their ground truth, A is more effective than B. Therefore, once B obtains a higher mutation score than A, a penalty will be counted when calculating Kendall for "mutation score". Although t can detect the specific fault, we cannot assume that another test case is less effective. It is possible another test case was used to detect another fault (e.g., a fixed fault in the previous version). Therefore, when considering the possibility of multiple defects (e.g., the intention of mutation testing), the assumption that A is more effective than B will be questioned.

In [34], researchers found that there is a strong correlation between mutant detection ratios and real fault detection ratios. Let A is the test suite that fails on the buggy version but passes on the fixed version, and B is the test suite that passes on the buggy and fixed versions. Meanwhile, A and B differ by one modified or added test. They assume that A is more effective than B. Their logic is that A can detect the specific defect, while B cannot. However, if the difference between A and B comes from "modified", the same threat to [41] will happen: the modification will be effective for the specific defect. Meanwhile, it cannot be denied that it is more effective for another possible defect before modification. If the difference comes from "added", the ground truth will be solid. As a result, we propose this ground truth in this paper.

*3.1.2 Proposed ground truth.* Based on Proposition 2 and Proposition 7, when comparing two test suites A and B, the following propositions (i.e., assumptions) are held:

if $B \subsetneq A$, then

$A$ is more effective than $B \Leftrightarrow FD(B) < FD(A)$  (1)

Here, *FD(T)* is the number of faults detected by T. Because B is the subset of A, we can conclude that A cannot be less effective than B. If A can detect more faults, it is the truth that A is more effective than B.

This is the ground truth adopted in this paper. Note that real faults may be expensive or hard to obtain in practice. As a result, in the experiment, we also compare the impact of using mutants instead of the real faults:

*Alternative ground truth.* Based on Proposition 2 and Proposition 5, the following propositions are held:

if $T_2 \subsetneq T_1$, then

$T_1$ is more effective than $T_2 \Leftrightarrow MS(M, T_2) < MS(M, T_1)$  (2)

Here, *M* is the whole possible mutant set, while *MS(M, T)* is the mutation score computed when executing *T* against *M*. Assume that *KM(M, T)* is the set of mutants in *M* killed by *T* and |*M*| is the number of mutants in *M*. Then, *MS(M, T) = |KM(M, T)|/|M|*. Note that the alternative ground truth is the baseline that is used for analyzing the impact of mutants.

After analyzing the threat of the existing ground truth and proposing our ground truth, we will introduce the approach to generate test suites for comparison in the next subsection.

## 3.2 Benchmark test suite generation

The approach to generate benchmark test suites depends on the ground truth. In many studies, researchers only compared test suites of the same size by their ground truth. This is to eliminate the confounding effect of size. For example, in many research, when researchers want to control the test suite size, it is popular to generate test suites with the same size by random sampling from an existing test pool. However, the question is that should we control the test suite size?

In [39], Chen et al. explained why test suite size is neither a confounding variable nor an independent variable that should be experimentally manipulated. Their opinion is that no tester will use "size" as the testing target, which we agree with. Besides, we would like to add that it may be difficult to compare test suites of the same size. For the simplest example, for two different test cases, t1 and t2. It is hard to conclude that t1 is more effective than t2 in most cases. The main reason is that they are likely to be designed for different test targets.

Therefore, let us start from the opposite of this problem: how to use the size for comparison? We decide to use different sizes to provide solid ground truth, as there is no basis for controlling the size to eliminate the confounding effect. It is a fact that adding test cases to a test suite does not decrease its effectiveness. As a result, when we generate a test suite with its subset, it is safe to compare them by real faults.

This approach is also in line with the actual testing scenario. In [39], to simulate the manual testing, Chen et al. "greedily select one test at a time to incrementally achieve adequacy for a given adequacy criterion" (i.e., MTE in our context). In essence, this is a process of continuous expansion of test suites, which is consistent with our approach.

Now we will describe how to generate test suites in this paper. At the high level, by the well-known benchmark Defects4J, for a buggy program, we use the developer-written tests with triggering test cases as the test suite A. Then, we filter all the triggering test cases to obtain B.

In detail, assume that for each fault, there is a buggy version and a fixed version and a test suite passed on fixed version, namely A, with at least one test case which fails only for the buggy version (i.e., triggering test cases). Then, we filter all the triggering test cases to obtain a test suite B. In other words, the set of triggering test cases is A-B. At this moment, we can assume that A is more effective than B in term of the specific defects. We hold this assumption not only for the buggy version but also for the fixed version. The reasons are two folds: for the buggy version, A can reveal the defects while B cannot; for the fixed version, the correctness is verified by the triggering tests cases.

Then the question is should we take more relationships among test suites into consideration? For example, for a test case t1 in B and a triggering test case t2, it is held that {t1, t2} is more effective than {t1}. This is also because that the former can reveal the defect while the latter cannot. Beside comparing A and B, should we take the comparison between {t1, t2} and {t1} into consideration? In our opinion, such a comparison is unnecessary, since this relationship is easier to be reflected. For example, in most cases, two test cases will achieve a higher line coverage than one of the two test cases. In essence, it is the increase of the size of the test suite leading to an increase of the coverage. For most criterion, it is easy to distinguish {t1, t2} with {t1} by the influence coming from size. To this concern, we need to eliminate the confounding effect as much as possible. Meanwhile, if a criterion can distinguish between A and B, for any subset of B, namely C, this criterion must be able to distinguish between C∪(A-B) and C. This is because A-B can uniquely contribute to the increase of the criterion metric. Thus, it is not necessary to take C into consideration.

As a result, we only compare the relationship between a complete set A and the subset B. This makes the relative size as similar as possible. For example, if there is only one triggering test case and 100 test cases in A. The size of A and B will differ by only 1%. If a criterion can accurately reflect the relationship between A and B, it will be sensitive enough to distinguish A and B's subsets. On the contrary, if a criterion cannot correctly reflect this relationship (e.g., A and B achieve the same line coverage), the set of triggering test cases is useless under the view of this criterion. Furthermore, it is hard to believe that this criterion has a strong distinguishing ability.

In a word, we only compare the effectiveness between A and B as the ground truth, which makes the relative size as similar as possible.

### 3.3 Agreement calculation

After generating the test suites for comparison, we need to choose a suitable indicator for the agreement. Ideally, the best test suite effectiveness metric will be consistent with the ground truth. Normally, we should use an indicator to report the agreement between the MTE and the ground truth.

The correlation analysis is a common approach to evaluate the MTE. We calculate the score provided by the MTE for each pair of (A, B) and compute the correlation coefficient (e.g., Spearman correlation, Pearson correlation, or Kendall correlation) between the MTE order and the ground truth order. In this paper, we use a variant of Kendall coefficient as the indicator to calculate the agreement. Specifically, in [38], Zhang et al. proposed an indicator OP (Order Preservation) to quantify the "order-preserving ability" of mutant reduction strategies. Given a mutation reduction strategy, OP measures the extent to which the mutation score orders among test suites are maintained before and after mutation reduction. In this paper, we adapt OP to quantify the ability of a test effectiveness metric in maintaining the ground truth order among test suites.

According to a ground truth criterion, assume we have generated $p$ test suite pairs with the order by $p$ faults. In our context, we assume that there is only one defect in a buggy version. This assumption is consistent with the existing benchmark (e.g., Defects4j). We use ($x$, $y$, operator) to represent a pair of test suite with order. $x$ and $y$ are the pair of test suites while operator is "more effective than" if $x$ is more effective than $y$ and is "as effective as" if $x$ is as effective as $y$ under a specific metric. We check whether each ($x$, $y$, operator) is satisfied or not under a test suite effectiveness metric $m$:

$$check(x, y, operator, m) = \begin{cases} 1 & if\ (x, y, operator)\ is\ held\ by\ m \\ 0 & else \end{cases}$$

Table 2. The information of projects in Defects4J

| Identifier | Project name | Number of bugs | Active bug ids | Used bug ids |
|---|---|---|---|---|
| Chart | jfreechart | 26 | 1-26 | 3,5-9,11-13, 16-18,20-25 |
| Cli | commons-cli | 39 | 1-5,7-40 | 1-5,7-40 |
| Closure | closure-compiler | 174 | 1-62,64-92,94-176 | 61-62,76,83,85,87,88, 91,92,94,95,97-106, 151-153,159,163 |
| Codec | commons-codec | 18 | 1-18 | 7-18 |
| Collections | commons-collections | 4 | 25-28 | - |
| Compress | commons-compress | 47 | 1-47 | 1-39 |
| Csv | commons-csv | 16 | 1-16 | 1-13 |
| Gson | gson | 18 | 1-18 | 1-13 |
| JacksonCore | jackson-core | 26 | 1-26 | 1-26 |
| JacksonDatabind | jackson-databind | 112 | 1-112 | 1-7,9-16,18,20,28,29, 34,35,39,42,43, 45,46,49,62,71,72 |
| JacksonXml | jackson-dataformat-xml | 6 | 1-6 | 1-6 |
| Jsoup | jsoup | 93 | 1-93 | - |
| JxPath | commons-jxpath | 22 | 1-22 | 1-22 |
| Lang | commons-lang | 64 | 1,3-65 | 1,3-65 |
| Math | commons-math | 106 | 1-106 | 1-14 |
| Mockito | mockito | 38 | 1-38 | 24,25,27-38 |
| Time | joda-time | 26 | 1-20,22-27 | 1-12,14-20,22-27 |

Furthermore, we define:

$W = \{(x, y, \text{operator}) \mid \text{check}(x, y, \text{operator}, m) = 1\}$

As such, we use the Order Preservation (OP) to measure the difference between the ground truth order by a ground truth criterion and the numerical MTE order by the metric $m$ as follows:

$$OP(m) = |W|/p$$

It is easy to know that OP has a value ranging from 0 to 1. The higher OP is, the smaller difference between the ground truth order and the MTE effectiveness order. If a test effectiveness metric has a small difference, it can be regarded as a good metric.

## 4 Experimental setup

In this section, we describe in detail the experimental design. First, we report the projects and tools used in our study. Then, we introduce the collection of the investigated test suite effectiveness metrics. Finally, we list the research question.

### 4.1 Subject programs and tools

*Projects.* We use Defects4J 2.0.0 to compare test effectiveness metrics, which is a benchmark with large real-world Java projects. There are reproducible real faults mined from source code repositories in the benchmark. For each fault, the benchmark contains a buggy version, a fixed version, and at least one triggering test cases. We use all the active bugs to compare the metrics. Among all bugs, we filter out those that fail the mutation testing or exceed the time limit (30 minutes). The remaining buggy versions are reported in Table 2.

*Mutants.* The tool we use to generate and run the mutants is Major [46], which is a widely used tool for Java mutation testing. To generate the whole possible mutant set, all mutators are used in this paper. After the mutants are executed, not only the mutation score but also the detail of which tests are executed can be obtained by the modified scripts accompanying Defects4J. It is worth noting that the mutation testing tool does not support failed test cases. To this concern, all the test criteria including coverage criteria are only implied for the fixed version.

*Tests.* For each fixed version, the manually written test cases, which are for the classes that were modified to fix the corresponding buggy version provided by Defects4J are used as the whole test suite.

*Coverage.* The tool we use to collect the coverage information is Cobertura [37], which is a widely used coverage tool. Statement coverage (SC) and branch coverage (BC) can be obtained by scripts accompanying Defects4J.

### 4.2 Test suite effectiveness metrics

In this study, we compare the accuracy of seven test suite effectiveness metrics: two code coverage metrics (SC and BC, which can be collected by Cobertura) and five mutation score metrics (the original mutation score MS, COS, RMS, SMS, and CMS). In our study, MS is obtained by Major directly.

*COS.* The 5-Operators Selection proposed by Offutt et al. [12] is the approach chosen in this paper. The 5 operators include the relational, logical, arithmetic, absolute, and unary insertion operators. For Major, we use 'LVR', 'AOR', 'ROR', 'LOR', and 'ORU' among all mutators as the implementation of COS. Then the mutation score computed on the mutants generated by these 5 mutators is used as the metric for measuring the effectiveness of test suites.

*RMS.* To implement RMS, given the ratio $n\%$ of selected mutants, we keep randomly selecting one of the remaining mutants with equal probability until there are $n\%$ mutants selected among all mutants. Finally, the mutation score computed by these mutants is the metric provided by RMS. In this paper, n=30.

*SMS.* To select the subsuming mutants for computing mutation scores, we use the matrix which records all test cases that kill a mutant. Then we select all the mutants, which are not subsumed by other mutants, to compute a mutation score as the metric.

*CMS.* To implement CMS, we first cluster all the mutants into different clusters and then select one mutant from each cluster to compute mutation scores. Before clustering, we need to collect feature data for each mutant. In this paper, an instance for a mutant is a 0-1 list based on the matrix in 4.2.4. For example, the instance of m is [1, 0, 1, 0], which means that the first and third test case kills m and the other two test cases do not kill m. Then, the k-means algorithm is applied to all mutant instances to classify the mutants into different clusters. Here, we set the "k" equaling to the number of selected mutants by SMS. Among each cluster, the mutants are guaranteed to be killed by similar test cases. After that, the selection is obtained by randomly selecting one mutant from each cluster. Finally, the mutation score computed by these mutants is the metric provided by CMS.

### 4.3 Research question

In this paper, we propose the following questions to evaluate the investigated metrics:

**RQ1: Based on ASSENT, which MTE is the most accurate metric under the ground truth criteria?**

The answer to RQ1 can find the best metric for evaluating test suite effectiveness. To investigate RQ1, for a MTE $m$, we compute the OP($m$) for a project with all the used bugs. For example, there are 6 bugs for the project CSV. Therefore, we can obtain 6 test suite pairs. In one pair, one test suite is the whole test suite and the other is its maximal subset without triggering test cases. By the ground truth, the former one is more effective than the latter. For a metric, we check the ground truth for 6 test suite pairs and compute OP for this metric. The larger the OP value is, the more accurate the metric $m$ is. Considering that there is randomness in 2 metrics (i.e., RMS and CMS), we will repeat each technology 20 times to obtain an average OP value.

Table 3. The average OP of seven metrics on 15 projects

| Project | MS | COS | RMS | SMS | CMS | SC | BC |
|---|---|---|---|---|---|---|---|
| Chart | 0.889 | 0.778 | 0.691 | 0.889 | 0.589 | 0.611 | 0.611 |
| Cli | 0.718 | 0.487 | 0.523 | 0.718 | 0.290 | 0.487 | 0.641 |
| Closure | 0.885 | 0.654 | 0.581 | 0.885 | 0.326 | 0.423 | 0.577 |
| Codec | 0.833 | 0.750 | 0.708 | 0.833 | 0.488 | 0.416 | 0.583 |
| Compress | 0.846 | 0.718 | 0.784 | 0.846 | 0.381 | 0.564 | 0.564 |
| Csv | 0.923 | 0.538 | 0.369 | 0.923 | 0.527 | 0.385 | 0.615 |
| Gson | 0.769 | 0.462 | 0.503 | 0.769 | 0.404 | 0.692 | 0.769 |
| Jacksoncore | 0.885 | 0.846 | 0.787 | 0.885 | 0.521 | 0.808 | 0.808 |
| JacksonDatabind | 0.800 | 0.567 | 0.450 | 0.800 | 0.300 | 0.733 | 0.733 |
| JacksonXml | 0.833 | 0.667 | 0.633 | 0.833 | 0.200 | 0.667 | 0.667 |
| Jxpath | 0.773 | 0.636 | 0.545 | 0.773 | 0.452 | 0.364 | 0.500 |
| Lang | 0.922 | 0.859 | 0.749 | 0.922 | 0.618 | 0.750 | 0.813 |
| Math | 0.571 | 0.428 | 0.418 | 0.571 | 0.443 | 0.500 | 0.357 |
| Mockito | 0.615 | 0.462 | 0.335 | 0.615 | 0.408 | 0.231 | 0.462 |
| Time | 0.760 | 0.720 | 0.618 | 0.760 | 0.410 | 0.360 | 0.440 |
| avg. | 0.801 | 0.638 | 0.580 | 0.801 | 0.424 | 0.533 | 0.609 |

**Table 4. The p-value (above '-') computed by the Wilcoxon signed rank test for 4 metrics after Benjamini-Hochberg adjustment and the cliff's delta δ (below '-') for 4 metrics.** (p <0.05) indicates there is a significant difference between the two corresponding metrics (i.e. bold value). (|δ| >0.147) indicates a nontrivial effective size (i.e. bold value).

| effect size\ p | COS | RMS | SC | BC |
|---|---|---|---|---|
| COS | - | **0.042** | 0.086 | 0.505 |
| RMS | -0.227(small) | - | 0.495 | 0.561 |
| SC | **-0.364(medium)** | **-0.173(small)** | - | 0.075 |
| BC | -0.106 | 0.111 | **0.271(small)** | - |

To see the difference among metrics, we list the average results for 15 projects.

**RQ2: Based on ASSENT, will using mutants instead of real faults lead to overestimation or underestimation of the MTE?**

In existing studies, the use of mutants instead of real defects is often a threat. However, it is unknown that will it lead to overestimation or underestimation of the MTE? To investigate RQ2, we use the alternative ground truth (in Section 3.1.2) to reevaluate the MTEs. In other words, OP is the proxy of the agreement between MS and a MTE. After that, we compare the results to RQ1. Finally, we can quantitatively analyze the differences between the two RQs.

## 5 Experimental results

### 5.1 Evaluation of MTEs based on ASSENT

Table 3 shows the average OP values of 7 metrics against 15 projects. Before analyzing the results, we need to reiterate that our focus is on the agreement between a metric and a ground truth criterion. As a result, we may not conclude that a metric is more practical than another metric by OP. RMS, for instance, can set different ratios for selection. Assume that RMS(n%) selects the n% of the mutants. The intuition is that the larger the selection is, the smaller the difference between the metric and the ground truth criterion is, while the larger the executing cost is. Due to this, it is unreasonable to conclude that RMS(90%) is more practical than RMS(30%) by OP. What we can conclude is that RMS(90%) can provide a more accurate metric on evaluating the effectiveness of test suites than RMS(30%). In order to facilitate readers to have an in-depth comparison of these metrics, we list their costs by the number of executed mutants. For SC and BC, none of the mutants need to be executed. For MS, SMS, and CMS, all of the mutants need to be executed. For COS, more than 30% of the mutants need to be executed on most programs. For RMS, 30% of the mutants need to be executed.

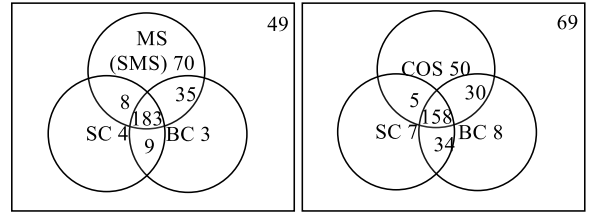

Figure 2. The Venn diagram to show the fault coupling between the MTEs.

**Table 5. The average OP of six metrics on 15 projects under alternative ground truth with comparison to RQ1.** The data in brackets indicates how much it has changed compared with RQ1.

| Project | COS | RMS | SMS | CMS | SC | BC |
|---|---|---|---|---|---|---|
| Chart | 0.889(+14%) | 0.803(+16%) | 1.000(+12%) | 0.700(+18%) | 0.722(+18%) | 0.722(+18%) |
| Cli | 0.770(+58%) | 0.805(+54%) | 1.000(+39%) | 0.572(+97%) | 0.615(+26%) | 0.718(+12%) |
| Closure | 0.770(+18%) | 0.697(+20%) | 1.000(+13%) | 0.443(+35%) | 0.539(+27%) | 0.615(+6%) |
| Codec | 0.917(+22%) | 0.875(+24%) | 1.000(+20%) | 0.655(+34%) | 0.593(+43%) | 0.750(+28%) |
| Compress | 0.872(+21%) | 0.812(+4%) | 1.000(+18%) | 0.535(+40%) | 0.615(+9%) | 0.615(+9%) |
| Csv | 0.615(+14%) | 0.446(+21%) | 1.000(+8%) | 0.704(+34%) | 0.308(-20%) | 0.693(+13%) |
| Gson | 0.693(+50%) | 0.735(+46%) | 1.000(+30%) | 0.635(+57%) | 0.770(+11%) | 0.857(+11%) |
| Jacksoncore | 0.961(+14%) | 0.902(+15%) | 1.000(+13%) | 0.637(+22%) | 0.770(-5%) | 0.846(+5%) |
| JacksonDatabind | 0.767(+35%) | 0.650(+44%) | 1.000(+25%) | 0.500(+67%) | 0.800(+9%) | 0.867(+18%) |
| JacksonXml | 0.834(+25%) | 0.800(+26%) | 1.000(+20%) | 0.367(+84%) | 0.834(+25%) | 0.834(+25%) |
| Jxpath | 0.864(+36%) | 0.773(+70%) | 1.000(+29%) | 0.680(+50%) | 0.500(+37%) | 0.637(+27%) |
| Lang | 0.937(+9%) | 0.837(+12%) | 1.000(+8%) | 0.696(+13%) | 0.828(+10%) | 0.891(+10%) |
| Math | 0.853(+99%) | 0.846(+103%) | 1.000(+75%) | 0.871(+96%) | 0.929(+86%) | 0.786(+120%) |
| Mockito | 0.846(+83%) | 0.720(+115%) | 1.000(+62%) | 0.793(+94%) | 0.615(+160%) | 0.846(+83%) |
| Time | 0.996(+38%) | 0.858(+38%) | 1.000(+31%) | 0.650(+58%) | 0.520(+44%) | 0.680(+54%) |
| avg. | 0.839(+32%) | 0.771(+33%) | 1.000(+24%) | 0.630(+48%) | 0.664(+24%) | 0.757(+24%) |

**Table 6. The p-value (above '-') computed by the Wilcoxon signed rank test for 6 metrics' change rates after Benjamini-Hochberg adjustment and the cliff's delta δ (below '-') for 6 metrics' change rates.** (p <0.05) indicates there are significant differences between the two corresponding metrics (i.e. bold value). (|δ| >0.147) indicates a nontrivial effective size (i.e. bold value).

| effect size\ p | COS | RMS | SMS | CMS | SC | BC |
|---|---|---|---|---|---|---|
| COS | - | 0.218 | **0.005** | **0.005** | 0.474 | 0.372 |
| RMS | 0.084 | - | **0.015** | 0.094 | 0.382 | 0.186 |
| SMS | **-0.236(small)** | **-0.240(small)** | - | **0.005** | 1 | 0.999 |
| CMS | **0.356(medium)** | **0.280(small)** | **0.596(large)** | - | 0.060 | **0.018** |
| SC | **-0.182(small)** | **-0.244(small)** | -0.04 | **-0.489(large)** | - | 0.877 |
| BC | **-0.311(small)** | **-0.298(small)** | **-0.173(small)** | **-0.587(large)** | -0.031 | - |

From Table 3, we have the following observations:

- MS and SMS are the most accurate metrics among the 7 metrics. On average, 80% of the enhancement of the test suite effectiveness will be expressed by MS or SMS. In other words, if we add triggering tests to detect an exposed fault, there is an 80% chance that the mutation score will increase. This observation is consistent with [34]. In their research, the OP for MS is 75%.
- The performance of SMS is the same as that of MS. First, we explain why SMS will increase, if MS increases. If a triggering test case leads to the increase of mutation score, it must kill a mutant which is not killed by other test cases. As a result, this mutant must be a subsuming mutant. Then, this mutant will be selected by SMS. Finally, this mutant leads to the increase of subsuming mutation score. Second, we explain why SMS will not increase, if MS does not increase. If a triggering test case does not lead to the increase of mutation score, it means the test suites A and B are not distinguished by the whole mutant set. For a subset of the whole set, it must not distinguish between A and B. Based on the two explanations, we can conclude that the performance of SMS is the same as that of MS.
- By common reduction strategies (COS, RMS), they cause a significant decline in OP (even the performance of BC and SC is comparable to them). To summarize, COS, RMS, CMS, BC, and SC may be risky and insensitive to evaluate test suite effectiveness. By adding triggering test cases, the effectiveness of test suits increases. However, such fact cannot be accurately reflected by these metrics.

Considering that the performance of COS, RMS, SC, and BC are close, we conduct a hypothesis test on whether there are significant differences between the four metrics by the Wilcoxon signed rank test. The p-value after Benjamini-Hochberg adjustment is shown in Table 4. To measure the magnitude of the difference, we employ the Cliff's delta δ, which is used for median comparison on effect size, to examine whether the magnitude of the difference is practically important. The delta is shown in Table 4. From Table 4, to summarize the results of p and delta, by OP, it is hard to conclude that COS is better than BC.

After comparing the OP values of several metrics, we also want to investigate the overlapping relationship between the maintained ground truth order. When the order among test suite pairs is not maintained by a MTE, it means this MTE does not take the corresponding fault into consideration, i.e., adding or deleting the triggering test cases will not change the metric value. In the following, if this happens, we say this fault is not considered by this MTE.

Figure 2 shows the Venn diagram for all the 361 faults. The number of faults considered by a MTE is attached in the corresponding circle. The overlapping part represents the faults considered by multiple metrics. Considering that there is randomness in RMS and CMS, we do not draw the diagram for them.

From the figure, we have the following observations:

- By all the metrics, 49 out of 361 faults are not considered. In other words, for these 49 faults, when testers add test cases into an existing test suite to detect them, none of the metric value will increase. This observation hints that there is a large space to improve the existing metrics or propose new metrics.
- When comparing MS to BC and SC, most faults are considered by all of them. Less than 5% ((4+9+3)/361) of faults are considered by coverage metrics uniquely. To this concern, their effectiveness is lower than MS.
- When comparing COS to BC and SC, coverage metrics are competitive. COS reflects 50 unique faults while coverage metrics reflect 49 (i.e., 7+8+34) unique faults. As a result, it is unreasonable to use COS instead of MS to compare coverage metrics. This observation proposes an empirical threat to [30].

To conclude, MS and SMS provide the most accurate metric among the 7 metrics under proposed ground truth.

### 5.2 Difference between the proposed ground truth and the alternative ground truth

Table 5 shows the average OP values of 6 metrics against 15 projects under the alternative ground truth. The data in brackets indicates how much it has changed compared with RQ1. For example, the "+14%" in the first cell is computed by (0.889-0.778)/0.778.

From the table, we have the following observations:
- In most cases, 6 metrics are overestimated by the alternative ground truth. Only SC may be underestimated in few cases. On average, all the metrics are overestimated by more than 20%.
- Mutation score metrics are more overestimated than code coverage metrics. In fact, if all the metrics are overestimated by the same ratio, the rank for the metrics will be the same. As a result, we need to analyze the influence of overestimation for the rank. In Table 3, the rank for COS, RMS, SMS, CMS, SC and BC is: 2, 4, 1, 6, 5, 3 while in Table 5, the rank is: 2, 3, 1, 6, 5, 4. The rank of BC changed most. As a result, it is not suitable to compare mutation score metrics to code coverage metrics under alternative ground truth. However, if we just compare mutation score metrics or code coverage metrics, the rank for the MTEs is consistent to the rank in RQ1.

Furthermore, for the change rate, we conduct a hypothesis test on whether there are differences between the 6 metrics' change rates by the Wilcoxon signed rank test. To measure the magnitude of the difference, the Cliff's delta is shown in Table 6. From Table 6, to summarize the results of p and delta, when using the alternative ground truth, CMS will be overestimated more than the other metrics.

To conclude, on average, all the metrics are overestimated by more than 20% by the alternative ground truth. Meanwhile, mutation score metrics are more overestimated than code coverage metrics on average. As a result, it is not suitable to compare mutation score metrics with code coverage metrics under the alternative ground truth.

## 6 Discussion

In discussion, we replicate our study across other programs, test suites, and mutation tools. As shown in Table 7, we select 10 programs from Apache [32] for study, as they were widely used in previous work [18, 27, 33, 34]. They range in size from dozens to thousands of lines. Note that without the information of real faults, we just compare 6 metrics by the alternative ground truth. The tool we use to generate and run the mutants is PIT 1.4.3 [35], which is a widely used tool for Java mutation testing. To generate the whole possible mutant set, "ALL" 29 mutators are used in this paper [36]. For PIT, we select *Conditionals Boundary Mutator*, *Increments Mutator*, *Invert Negatives Mutator*, *Math Mutator*, *Negate Conditionals Mutator*, *Remove Conditionals Mutator*, *Experiment Switch Mutator*, *Negation Mutator*, *Arithmetic Operator Replacement Mutator*, *Bitwise Operator Mutator*, *Relational Operator Replacement Mutator* and *Unary Operator Insertion Mutator* among all 29 mutators as the implementation of COS. To generate A and B, we randomly select a test suite with $k$ ($k>1$) test cases and one of its subsets with $k$-1 test cases to obtain a pair. We repeat the selecting process 100 times to obtain 100 pairs of test suites for a program.

Table 8 shows the average OP values of 6 metrics under alternative ground truth. From the table, we have the following observations:

- The approach to generating test suites will influence the results greatly. RMS achieves the OP value more than 0.9 in most of the cases. This is greatly different from the result in RQ2. This observation leads us to suggest that research should pay more attention to the possible threat by the test suite generation. When using the alternative ground truth, the way to generate the test suite in RQ1 may not be suitable.[1]
- SMS will not achieve the highest (i.e., 1.0) OP by more test suite pairs. We have explained why SMS is the same as MS when comparing A and B in RQ1. However, when more test suite pairs are taken into consideration, especially the test suite pairs which cannot kill the subsuming mutants, the OP of SMS will decrease.

To conclude, not only the ground truth, but also the generation of test suites will influence the conclusion. To this concern, a suitable way to generate test suites according to the used ground truth must be defined carefully.

## 7 Threats to validity

The first threat comes from the proposed ground truth. Because we adopt a conservative ground truth, whose advantage is easy to be recognized by researchers, it makes our conclusion is conservative. In other words, a MTE with a low OP is useless, while a MTE with a high OP may not be useful. For example, if we use the size of test suites as a metric, by our ground truth, it will achieve the 1.0 (i.e., highest) OP. As a result, we cannot deny that using the best metric under our ground truth may be risky in other cases. However, the risk of a MTE revealed by our study is objective. For a MTE with a low OP, it cannot be sensitively aware of the improvement of the test suite effectiveness. As a result, it has a low ability to evaluate test suite effectiveness.

Table 7. The information of programs in discussion

| | Program | LOCs | Path | Tests | Mutants | MS |
|---|---|---|---|---|---|---|
| J1 | Crypt | 25 | org.apache.commons.codec.digest | 12 | 63 | 0.952 |
| J2 | StringUtils | 48 | org.apache.commons.codec.binary | 234 | 189 | 0.894 |
| J3 | Md5Crypt | 107 | org.apache.commons.codec.digest | 16 | 736 | 0.836 |
| J4 | DefaultParser | 178 | org.apache.commons.cli | 64 | 1361 | 0.796 |
| J5 | Option | 192 | org.apache.commons.cli | 303 | 829 | 0.668 |
| J6 | ArithmeticUtils | 207 | org.apache.commons.math3.util | 323 | 3166 | 0.714 |
| J7 | ResizableDoubleArray | 217 | org.apache.commons.math3.util | 60 | 2126 | 0.733 |
| J8 | UnixCrypt | 311 | org.apache.commons.codec.digest | 9 | 2998 | 0.981 |
| J9 | HelpFormatter | 416 | org.apache.commons.cli | 33 | 1543 | 0.807 |
| J10 | Dfp | 1702 | org.apache.commons.math3.dfp | 193 | 15134 | 0.936 |

Table 8. The average OP of several metrics on 10 programs under alternative ground truth

| Program | COS | SMS | CMS | RMS | SC | BC |
|---|---|---|---|---|---|---|
| J1 | 0.823 | 0.903 | 0.867 | 0.920 | 0.662 | 0.609 |
| J2 | 0.952 | 0.977 | 0.960 | 0.972 | 0.846 | 0.871 |
| J3 | 0.728 | 0.554 | 0.635 | 0.948 | 0.553 | 0.482 |
| J4 | 0.943 | 0.726 | 0.687 | 0.929 | 0.587 | 0.535 |
| J5 | 0.961 | 0.915 | 0.906 | 0.964 | 0.826 | 0.830 |
| J6 | 0.993 | 0.909 | 0.903 | 0.985 | 0.835 | 0.832 |
| J7 | 0.983 | 0.731 | 0.752 | 0.971 | 0.566 | 0.583 |
| J8 | 0.817 | 0.509 | 0.529 | 0.860 | 0.593 | 0.449 |
| J9 | 0.926 | 0.609 | 0.469 | 0.906 | 0.507 | 0.513 |
| J10 | 0.892 | 0.882 | 0.837 | 0.958 | 0.576 | 0.618 |
| avg. | 0.902 | 0.772 | 0.755 | 0.941 | 0.655 | 0.632 |

The second threat is that our evaluation is performed on a limited set of Java projects. To mitigate this threat, we replicate our study across more programs, test suites, and mutation tools in the discussion.

## 8 Conclusion

---
[1] Note that the aim of RQ2 is to show the impact of using mutants instead of real faults. As a result, the "test suite" is consistent to RQ1.

We propose a framework ASSENT for evaluating the accuracy of any test suite effectiveness metric. Under this framework, the accuracy of test suite effectiveness metrics can be consistently and accurately evaluated and compared. We apply this framework to compare representative test effectiveness metrics. As a result, we find that: based on the real faults, mutation score and subsuming mutation score are the best metrics; if we use mutants as the alternative ground truth, MTEs will be overestimated by more than 20% in values. In the future, we strongly suggest that any newly proposed test effectiveness metrics should be evaluated under ASSENT to demonstrate the practical value.

## REFERENCES

[1] I. Ahmed, R. Gopinath, C. Brindescu, A. Groce, C. Jensen. Can testedness be effectively measured? FSE 2016: 547-558.
[2] X. Cai. Coverage-based testing strategies and reliability modeling for fault-tolerant software systems. Ph.D. dissertation, Chinese Univ. Hong Kong, Hong Kong, 2006.
[3] X. Cai, M. R. Lyu. The effect of code coverage on fault detection under different testing profiles. ACM SIGSOFT Software Engineering Notes, 30(4), 2005: 1-7
[4] P. S. Kochhar, F. Thung, D. Lo. Code coverage and test suite effectiveness: Empirical study with real bugs in large systems. SANER 2015: 560-564.
[5] R. G. Hamlet. Testing programs with the aid of a compiler. IEEE Transactions on Software Engineering, 3(4), 1977: 279-290
[6] Y. Jia, M. Harman. An analysis and survey of the development of mutation testing. IEEE Transactions on Software Engineering, 37(5), 2011: 649-678.
[7] M. Papadakis, M. Kintis, J. Zhang, Y. Jia, Y. Traona, M. Harmanb. Mutation testing advances: An analysis and survey. Advance in Computers, 112, 2019: 275-378.
[8] B. Baudry, F. Fleurey, Y. L. Traon. Improving test suites for efficient fault localization. ICSE 2006: 82-91.
[9] T. Ball. A Theory of Predicate-Complete Test Coverage and Generation. Formal Methods for Components and Objects. 2004: 1-22
[10] J. R. Larus. Whole Program Paths. Programming Language Design and Implementation.1999: 259– 269.
[11] A. S. Namin, J. H. Andrews, D. J. Murdoch. Sufficient mutation operators for measuring testing effectiveness. ICSE 2008: 10-18.
[12] J. Offutt, A. Lee, G. Rothermel, R. H. Untch, C. Zapf. An experimental determination of sufficient mutant operators. ACM Transactions on Software Engineering and Methodology, 5(2), 1996: 99-118.
[13] P. Delgado-Prez, S. Segura, I. Medina-Bulo. Assessment of C++ object-oriented mutation operators: A selective mutation approach. Software Testing, Verification and Reliability, 27(4-5), 2017: n/a-n/a.
[14] M. E. Delamaro, J. Offutt, P. Ammann. Designing deletion mutation operators. ICST 2014:11-20.
[15] M. E. Delamaro, L. Deng, V. H. S. Durelli, N. Li, J. Offutt. Experimental evaluation of SDL and one-op mutation for C. ICST 2014: 203-212.
[16] A. P. Mathur. Performance, effectiveness, and reliability issues in software testing. COMPSAC,1991:604-605.
[17] B. Kurtz, P. Ammann, J. Offutt, M. E. Delamaro, M Kurtz, N. Gokce. Analyzing the validity of Selective mutation with dominator mutants. FSE 2016:571-582.
[18] D. Gong, G. Zhang, X. Yao, F. Meng. Mutant reduction based on dominance relation for weak mutation testing. Information and Software Technology. 81, 2017: 82-96.
[19] R. A. DeMillo, R. J. Lipton, F. G. Sayward. Hints on test data selection: help for the practicing programmer. IEEE Computer, 11(4), 1978: 34-41.
[20] B. Kurtz, P. Ammann, M. E. Delamaro, J. Offutt, L. Deng. Mutant subsumption graphs. ICSTW 2014: 176–185.
[21] S. Hussain. Mutation clustering．M.S. thesis London, UK King's College. 2008.
[22] M. Papadakis, N. Malevris. An empirical evaluation of the first and second order mutation testing strategies. ICST 2010:90-99.
[23] W. E. Wong, A. P. Mathur. Reducing the cost of mutation testing：an empirical study. Journal of Systems and Software. 31(3), 1995: 185-196.
[24] B. Kurtz, P. Ammann, J. Offutt. Static analysis of mutant subsumption. ICST 2015: 1-10.
[25] K. Jalbert, J.S. Bradbury. Predicting mutation score using source code and test suite metrics. ICSE 2012: 42-46.
[26] K. Jalbert. Predicting mutation score using source code and test suite metrics. M.S. thesis, University of Ontario Institute of Technology, 2012.
[27] J. Zhang, L. Zhang, M. Harman, D. Hao, Y. Jia, L. Zhang, Predictive mutation testing. IEEE Transactions on Software Engineering, 45(9), 2019: 898-918.
[28] P. Zhang, Y. Li, W. Ma, Y. Yang, L. Chen, H. Lu, Y. Zhou, B. Xu. CBUA: A probabilistic, predictive, and practical approach for evaluating test suite effectiveness. IEEE Transactions on Software Engineering, 2020. https://doi.org/10.1109/TSE.2020.3010361
[29] G. Grano, F. Palomba, H.C. Gall. Lightweight assessment of test-case effectiveness using source-code-quality indicators. IEEE Transactions on Software Engineering, accepted, 2019.
[30] M. Gligoric, A. Groce, C. Zhang, R. Sharma, M. A. Alipour, D. Marinov. Guidelines for Coverage-Based Comparisons of Non-Adequate Test Suites. ACM Transactions on Software Engineering and Methodology, 24(4), 2015: 22:1-22:33.
[31] R. Gopinath, C. Jensen, A. Groce. Code coverage for suite evaluation by developers. ICSE 2014: 72-82.
[32] Apache. http://commons.apache.org
[33] L. Inozemtseva, R. Holmes. Coverage is not strongly correlated with test suite effectiveness. ICSE 2014: 435-445.
[34] R. Just, D. Jalali, L. Inozemtseva, M. D. Ernst, R. Holmes, G. Fraser. Are mutants a valid substitute for real faults in software testing? FSE 2014: 654-665.
[35] PIT. http://pitest.org/
[36] PIT mutators. https://pitest.org/quickstart/mutators
[37] Cobertura. https://github.com/cobertura
[38] P. Zhang, Y. Wang, X. Liu, Y. Li, Y. Yang, Z. Wang, X. Zhou, L. Chen, Y. Zhou. Mutant reduction evaluation: what is there and what is missing? ACM Transactions on Software Engineering and Methodology, accepted, 2021.
[39] Y. T. Chen, R. Gopinath, A. Tadakamalla, M. D. Ernst, R. Holmes, G. Fraser, P. Ammann, R. Just. Revisiting the Relationship Between Fault Detection, Test Adequacy Criteria, and Test Set Size. ASE 2020: 237-249.
[40] F. Hariri, A. Shi, V. Fernando, S. Mahmood, D. Marinov. Comparing Mutation Testing at the Levels of Source Code and Compiler Intermediate Representation. ICST 2019: 114-124.
[41] M. Papadakis, D. Shin, S. Yoo, D. Bae. Marinov. Are Mutation Scores Correlated with Real Fault Detection? ICSE 2018: 537-548.
[42] D. Shin, S. Yoo, D. Bae. Marinov. A Theoretical and Empirical Study of Diversity-Aware Mutation Adequacy Criterion. IEEE Transactions on Software Engineering, 2018: 914-931.
[43] T. T. Chekam, M. Papadakis, Y. L. Traon, M. Harman. An Empirical Study on Mutation, Statement and Branch Coverage Fault Revelation that Avoids the Unreliable Clean Program Assumption. ICSE 2017: 597-608.
[45] A. S. Namin, J. H. Andrews. The influence of size and coverage on test suite effectiveness. ISSTA 2009: 57-68
[46] R. Just, F. Schweiggert, G. M. Kapfhammer. MAJOR: An efficient and extensible tool for mutation analysis in a Java compiler. ASE 2011: 612-615